\documentclass[twocolumn]{aastex631}

\usepackage{enumitem}
\usepackage{amsmath}
\usepackage{mathdots}
\usepackage{multirow}
\usepackage{mathrsfs}
\usepackage{float}

\usepackage{bm}

\begin{document}

% \title{Efficient parameterization of ion velocity distribution functions using Slepian functions}
\title{Recovering Ion Distribution Functions: \\
I. Slepian Reconstruction of VDFs from MMS and Solar Orbiter}
\shorttitle{Slepian Reconstruction of Ion VDFs from MMS and SolO}

\correspondingauthor{Srijan Bharati Das}
\email{srijanbdas@alumni.princeton.edu}

\author[0000-0003-0896-7972]{Srijan Bharati Das}
\affiliation{Center for Astrophysics | Harvard \& Smithsonian, 
60 Garden Street, Cambridge, MA 02138, USA.}

\author[0000-0003-4747-6252]{Michael Terres}
\affiliation{Center for Astrophysics | Harvard \& Smithsonian, 
60 Garden Street, Cambridge, MA 02138, USA.}

\shortauthors{Bharati Das \& Terres}

% \author[0009-0006-8967-001X]{Laura Garavito}
% \affiliation{Center for Astrophysics | Harvard \& Smithsonian, 
% 60 Garden Street, Cambridge, MA 02138, USA.}
% \affiliation{Tufts University, 419 Boston Ave, Medford, MA 02155, USA.}

%% Note that the \and command from previous versions of AASTeX is now
%% depreciated in this version as it is no longer necessary. AASTeX 
%% automatically takes care of all commas and "and"s between authors names.

%% AASTeX 6.31 has the new \collaboration and \nocollaboration commands to
%% provide the collaboration status of a group of authors. These commands 
%% can be used either before or after the list of corresponding authors. The
%% argument for \collaboration is the collaboration identifier. Authors are
%% encouraged to surround collaboration identifiers with ()s. The 
%% \nocollaboration command takes no argument and exists to indicate that
%% the nearby authors are not part of surrounding collaborations.

%% Mark off the abstract in the ``abstract'' environment. 
\begin{abstract}

Plasma velocity distribution functions (VDFs) constitute a fundamental observation of numerous operational and future missions. An efficient parameterization of VDFs is crucial for (1) preserving enough information to investigate macroscopic moments along with kinetic effects, (2) producing smooth distributions whereby it is possible to perform derivatives in phase space to support numerical solvers, and (3) economic data management and its storage. Previous studies have used spherical harmonics as an efficient basis for representing electron VDFs. In this paper, we present a novel algorithm targeted towards decomposing ion VDFs measured by electrostatic analyzers onboard Magnetospheric Multiscale Mission (MMS) and Solar Orbiter (SolO) spacecrafts. We use Slepian functions, custom-designed bases providing compact support in phase space, initially developed in information theory and later used for terrestrial and planetary applications. In this paper, we choose well-studied, well-measured, and complex intervals from MMS and SolO containing a range of simpler gyrotropic and agyrotropic distributions to benchmark the robustness of our reconstruction method. We demonstrate the advantages of using Slepian functions over spherical harmonics for solar wind plasma distributions. We also demonstrate that our choice of basis representation efficiently preserves phase space complexities of a 3D agyrotropic distribution function.  
This algorithm shown in this study will be extended to Parker Solar Probe and future missions such as Helioswarm.
% will be extended to (a) single spacecraft partial coverage VDF parameterization and recovery (Parker Solar Probe --- Solar Probe ANalyzer \& Solar Probe Cup), and to (b) multi-spacecraft partial coverage VDF parameterization, cross-calibration and recovery (Helioswarm --- hub and nodes). 

\end{abstract}

%% Keywords should appear after the \end{abstract} command. 
%% The AAS Journals now uses Unified Astronomy Thesaurus concepts:
%% https://astrothesaurus.org
%% You will be asked to selected these concepts during the submission process
%% but this old "keyword" functionality is maintained in case authors want
%% to include these concepts in their preprints.
\keywords{plasmas --- methods: data analysis --- methods: analytical --- Sun: solar wind}

%% From the front matter, we move on to the body of the paper.
%% Sections are demarcated by \section and \subsection, respectively.
%% Observe the use of the LaTeX \label
%% command after the \subsection to give a symbolic KEY to the
%% subsection for cross-referencing in a \ref command.
%% You can use LaTeX's \ref and \label commands to keep track of
%% cross-references to sections, equations, tables, and figures.
%% That way, if you change the order of any elements, LaTeX will
%% automatically renumber them.
%%
%% We recommend that authors also use the natbib \citep
%% and \citet commands to identify citations.  The citations are
%% tied to the reference list via symbolic KEYs. The KEY corresponds
%% to the KEY in the \bibitem in the reference list below. 

% \begin{itemize}
%     \item Introduction
%     \item Methods
%     \begin{itemize}
%         \item Finding Axis of Gyrotopy [Fig 1]
%         \item Slepian Reconstruction [Fig 2, Fig 3]
%         \item In Preparation for FOV Restricted ESAs [Fig 4]
%     \end{itemize}
%     \item Final Remarks [Fig 5]
%     \item Appendix
%     \begin{itemize}
%         \item Math
%         \item Slepian on Polar Cap
%         \item Electron VDF with Spherical Harmonics: Comparison to Vinas et al. 2009.
%         \itme Ion VDF with Spherical Harmonics: Motivation for Slepian Functions
%     \end{itemize}
% \end{itemize}

\section{Introduction} \label{sec:intro}

Particle Velocity distribution functions (VDFs) are crucial for understanding plasma kinetic and macroscopic properties (i.e., density, velocity, and temperature). A VDF represents the probability density of finding particles at specific points in phase space. In the weakly collisional, multi-species solar wind, interparticle collisions occur infrequently, inhibiting the distributions from reaching local thermodynamic equilibrium (LTE) \citep{Marsch_2006, verscharen2019multi}. Ion VDFs which deviate from LTE have been observed to exhibit one or more of the following: temperature anisotropies ($T_{\perp}/T_{\parallel} \neq 1$) \citep{Kasper_2002, Huang_2020}, field-aligned beams \citep{Alterman_2018, Verniero_2020}, temperature differences between ion species \citep{Kasper_2008, Kasper_2017}, and drifts between multiple ion populations \citep{gershman2012solar, bourouaine2013limits}. Non-LTE characteristics serve as sources of free energy that interact with the fundamental electromagnetic fluctuations via wave-particle interactions \citep[see][for more details]{verscharen2019multi}. Ultimately, our capacity to comprehensively observe VDFs is crucial for understanding the fundamental kinetic processes driving plasma heating, acceleration, and turbulence.

Current spacecraft missions rely on Electrostatic Analyzers (ESAs) to measure the thermal (low energy) portion of the three-dimensional particle VDFs\footnote{It is important to note that ESAs do not provide a continuum of measurements, and only discrete points in phase space are sampled. Additionally, limitations of the angular extent and energy resolution cannot cover all of phase space. Hence, the measured distribution is not a \textit{complete} measure of phase space.}. 
The resultant data product is sparsely sampled and depends on the number of energy channels, azimuthal anodes, and selected elevation angles. Depending on the particle species, plasma temperature, and distance from the Sun where the in-situ observation is made, the VDFs span only a part of phase space. Most often, ions in the solar wind are cold and supersonic. This necessitates high angular and energy resolution to accurately quantify essential observational signatures, such as the proton heat flux \citep{Wilson_2022}.

Due to the coarse nature of the observations, bi-Maxwellian models are often implemented to provide a smooth and comprehensive representation of VDFs. The bi-Maxwellian is an extension of the standard Maxwellian distribution, representing the equilibrium state for highly collisional plasmas.
% is a non-thermal, multi-population extension of the standard Maxwellian distribution, which represents the equilibrium state for highly collisional plasmas. 
Bi-Maxwellians reasonably approximate ion VDFs in the solar wind while preserving non-LTE features like temperature anisotropy. Computationally, bi-Maxwellian fits facilitate straightforward calculations of phase space density gradients, enabling numerical solutions to linear dispersion relations \citep{verscharen2019multi}.

However, bi-Maxwellian models are inadequate to capture the phase space complexities observed in measured VDFs. Additionally, the convergence of iterative non-linear algorithms adopted for bi-Maxwellian fits is highly sensitive to the choice of initialization parameters \citep{Verniero_2020, Klein_2021, Bowen_etal_2023}. 
This problem may be alleviated by employing linear inversions, which by construction, converge in one step. 
Such fitting methods have been used previously for various applications in reconstructing particle VDFs. Of note, polynomial-based reconstructions were implemented, with notable choices being Hermite Polynomials \citep{Servidio_etal_2017, Bowen_etal_2023}, Legendre Polynomials \citep{carcaboso2020characterisation}, Spherical Harmonics \citep{Vinas_and_Gurgiolo_2009}, and recently Radial Basis functions \citep{Bowen_etal_2023}. However, the plasma population and the overall temperature dictate how well the various optimization procedures perform. 

Electron VDFs are more accessible to fit as they occupy all of phase space due to their relatively large thermal velocity $(m_e \ll m_i)$. Since ESAs measure phase space on a spherical grid at discrete energy levels, the parameterization of electron distribution in terms of spherical harmonics constitutes an efficient basis that accurately preserves anisotropies, asymmetries, and fluid moments \citep{Vinas_and_Gurgiolo_2009}. A limited number of coefficients effectively encapsulate the information, wherein the truncation of the series determines the degree of granularity at which the distribution is resolved \citep{Vinas_and_Gurgiolo_2009}. However, ion distributions (particularly in the solar wind) are localized to a small angular extent in velocity phase space due to smaller thermal speeds (i.e., relatively massive particles) than electrons. Decomposing ion VDFs in the solar wind into spherical harmonics is inefficient. The finite structures of the measured ion distributions require a considerable number of coefficients to resolve.

In this work, we address this limitation by using 
Slepian functions. These are a family of orthogonal basis functions having simultaneous finite support in spatial and spectral space, thereby useful to represent compact signals. Slepian functions may be locally concentrated in space and locally limited in spectra, or vice versa. 
% --- specialized basis functions designed to optimally span a desired patch on the surface of a sphere or a Cartesian plane. 
These basis functions, designed initially in information theory \citep{Slepian_1961, Landau_1961, Landau_1962, Slepian_1983} and successively applied in terrestrial geophysics and planetary sciences, have found applications in investigating geodesy of polar caps \citep[][]{Slepian_polar_caps}, magnetization of Australian lithosphere \citep[][]{Slepian_australia} or melting of ice-sheets on Greenland \citep[][]{Slepian_Greenland}. More recently, in the solar context, Slepian functions have been suggested as an efficient basis to infer solar internal magnetism \citep[][]{Das2022} and have been used to image the three-dimensional structure of an average solar supergranule \citep[][]{Hanson24_Nat}. 

In this study, we use Slepian functions localized to optimally span a polar cap defined on the surface of constant energy shells in phase space. This renders compact support for cold solar wind plasma distributions. Furthermore, this allows for a smooth representation of ion VDFs that otherwise suffer from singular or null bin counts due to discrete finite-resolution effects \citep{verscharen2019multi}.
% \textcolor{red}{Furthermore, finite-resolution effects, such as singular or null bin counts, may drastically alter the apparent interpretation of the particle distribution \citep{verscharen2019multi}.}\\
% This study is the first in a series of papers where we will use Slepian functions to parameterize ion VDFs 
% --- \textcolor{red}{building toward a more complete particle distribution in solar wind plasma}. 
We outline this method to optimally reconstruct ion VDFs from ESA measurements in both warm (magnetosheath) and cold (solar wind) plasmas. Furthermore, we argue that the Slepian Basis Reconstruction (SBR) method is an efficient data compression technique, especially for ion VDFs where spherical harmonics are unsuitable. The selected data is discussed in Section~\ref{sec:data}. Section~\ref{sec:methods} details the SBR method and discusses the moment comparison metric and visual features. 
% We then discuss the future applications and planned extensions of this technique to SPAN-i onboard Parker Solar Probe in Section~\ref{sec:methods}. 
Finally, in Section~\ref{sec:final_remarks}, we detail extensions of this work to PSP and SPAN-i and discuss further science applications. 

% (accounting for FOV restrictions imposed by spacecraft architecture --- such as a heat shield on-board PSP). \textit{Upcoming work-in-progress will lay down the recipe to combine ESA and FC measurements to constrain the VDF recovery as best as possible for missions where these instruments are onboard the same spacecraft. Ultimately we will extend this to inferring multi-point VDFs from combining measurements across ESA and FC located on multiple spacecrafts, such as in a constellation (future missions such as Helioswarm).} 

% \texttt{Add paragraph on the science relevance of this paper to multiple investigators using the different instruments we are using/priming in this paper --- Wind, MMS, SoLO, PSP and Helioswarm.}

% We believe that the applicability of this work will extend to various areas of space plasma. We show that the Slepian functions are optimal for representing cold, fast-moving plasmas as found in the solar wind.

\section{Data} \label{sec:data}
For this work, we investigate well-studied, well-measured, and complex ion distribution events to benchmark the SBR method. This study uses ion distribution measurements from the Magnetospheric Multiscale (MMS) and Solar Orbiter (SolO) missions. The ion VDFs represent the measured particle fluxes defined as a function of time, energy, elevation, and azimuth. 
% All data in this work is restructured to be in the format of time, energy, elevation, and azimuth. 
All methods and techniques are identical for each dataset. The temperature of the plasma populations is used to define the angular extent of our Slepian basis (see Section~\ref{sec:methods}). 
% with the exception of defining the angular extent of the Slepian basis (see Section~\ref{sec:methods}). 

MMS is a constellation of four identical spacecraft with the same instrument suite. The Fast Plasma Instrument (FPI) measures the ``full'' 3D ion velocity distribution in the near-Earth environment. FPI samples 32 energy channels, 32 azimuthal (anode) angles, and 16 elevation (deflector) angles at a cadence of 150 milliseconds during the burst mode \citep{MMS_FPI_2016}. For our study, we select the burst mode interval on 2016-01-11 from 00:57:04 to 01:00:33 with 1399 timestamps. This is the same interval analyzed by \cite{Servidio_etal_2017} where they represent the VDF in a 3D Hermite polynomial basis. For our purpose, we present observations from only one of the four probes. 

The Proton and Alpha Sensor (PAS) instrument, a part of the Solar Wind Analyzer suite (SWA) onboard SolO, provides 3D VDF measurements in the Solar Wind \citep{Owen_SWA_2020}. The electrostatic analyzer PAS measures VDF over 96 energy channels, 11 azimuthal (anode) angles, and 9 elevation (deflection) angles. Unlike MMS, PAS only measures a targeted subset of phase space spanning $-24^{o}$ to $+42^{o}$ in azimuth, $\pm22.5^{o}$ in elevation. Seeing as solar wind ions have a high Mach number, they generally occupy only a fraction of targeted phase space, suggesting that 32 energies and five elevations are sufficient to characterize the proton-alpha population \citep{Louarn_2021}. For our study, we select the interval on 2020-07-16TT18:15:00 to 18:45:00. This interval, which was also investigated by \cite{Lavraud_etal_2021}, was recorded by PAS near the Heliospheric current sheet.     

\section{Methodology} \label{sec:methods}
The fitting code architecture of SBR is developed with a data pre-processing pipeline which casts the data from MMS and SolO into the same ordering in dimensions before feeding into the fitting routine. However, an important distinction we make between MMS and SolO is that the MMS VDFs span a much larger region in azimuth-elevation (henceforth referred to as $\phi$ and $\theta$, respectively) compared to SolO where the VDFs are locally compact in $(\theta,\phi)$.  

\subsection{Slepian Basis Reconstruction} \label{sec:slepian_recon}
We use spatially-concentrated and spectrally-bandlimited Slepian functions in $(\theta,\phi)$ for representing distributions belonging to shells of constant energy. We allow these Slepian functions to be localized on a polar cap in velocity phase-space and truncated beyond a chosen angular degree $L_{\rm{max}}$. An example of such polar cap Slepians in shown in Appendix~\ref{appendix: slepian_background} where the polar cap extent is $85^{\circ}$ and an angular degree cutoff at $L_{\rm{max}} = 8$. The maximum allowable band-limit in spectral space is governed by the Nyquist wavenumber $L_{\rm{max, Nyq}}$. For MMS, this is calculated based on the grid spacing $\Delta \theta \sim 11.25^{\circ}$ where the grids in $\theta \sim [5.625^{\circ}, 174.375^{\circ}]$ are cell-centered measurements spanning bins ranging from $(0^{\circ}, 180^{\circ})$. So, the maximum number of complete wavelengths that can be captured with this grid resolutions is $L_{\rm{max, Nyq}}= 180 / (\Delta \theta)$. This gives us $L_{\rm{max, Nyq}} = 16$ for MMS-FPI. Since the Slepian functions on the surface of a sphere are constructed from linear combinations of spherical harmonics, the underlying spherical harmonics basis used to construct these Slepian functions is truncated beyond angular degree 16. 

Since SBR is developed with the goal of being able to custom fit every VDF depending on the thermal speed and the degree of gyrotropy (or agyrotropy), we need to specify a contour in velocity phase space for generating these Slepian functions. Since, for MMS and SolO we do not have significant field of view restrictions, we choose to perform the VDF fitting in $(\theta, \phi)$ space to be able to maximally retain non-thermal features of the distribution. Out of the different classes of Slepian basis functions that can be evaluated on the surface of a sphere, we use the Slepians on a polar cap. This is because generating Slepian functions on an axisymmetric domain (such as a polar cap) is computationally simpler. An outline of the application of Slepian functions on a polar cap is provided in Appendix~\ref{appendix: slepian_background}. We have highlighted only the salient features of the theoretical formalism which are immediately required to complement the description of our methodology. We refer the author to the original work of \cite{Slepian_polar_caps} for a complete description.

\subsection{Application to MMS-FPI ion distributions} \label{sec: MMS_REC}

\begin{figure*}
    \centering
    \includegraphics[width=\linewidth]{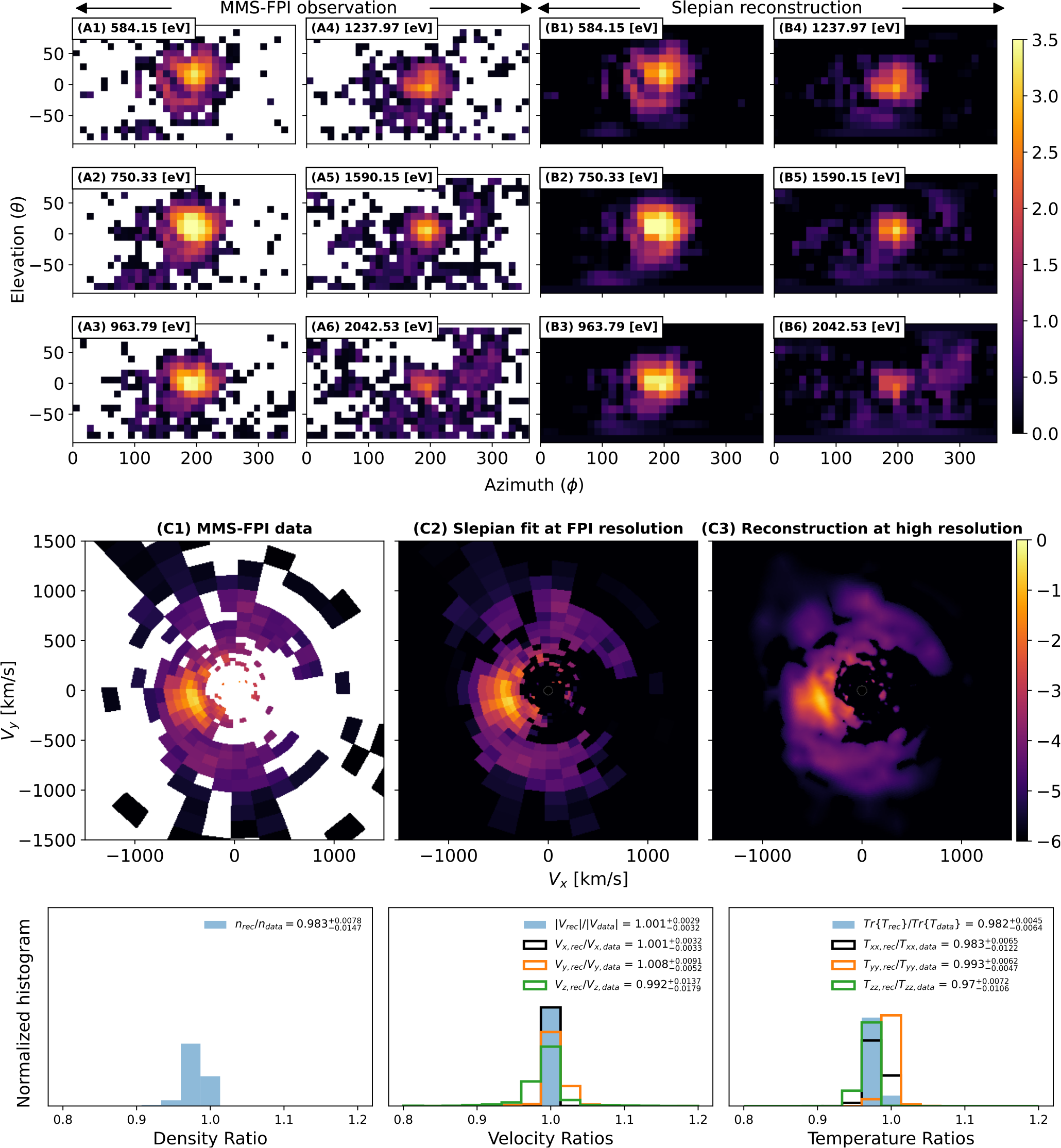}
    \caption{\textbf{MMS-FPI energy shell reconstruction plots.} Figures on the two left columns (A1)-(A6) show selected energy channels of MMS-FPI observation at time 2016-01-11/00:58:24 UTC. The corresponding energy shell values are written in the panel headers. The right two columns show the corresponding reconstructions in (B1)-(B6), respectively. Panels (C1)-(C3) shows the data and reconstructions on the closest Cartesian slice to the $V_x - V_y$ plane. (C1) shows the slice of MMS-FPI data at $\theta = 5.625^{\circ}$ off the $V_z = 0$ plane and (C2) shows the Slepian reconstruction at the same resolution as the FPI measurement. (C3) shows a high resolution reconstruction where we used a cubic B-spline interpolation in energy to ensure smoothness and a higher resolution Slepian function weighted by the same coefficients as inferred from the FPI resolution Slepian functions. The \textit{bottom} panel compares the density and velocity moments obtained from the data and reconstruction. Each histogram corresponds to the moment ratios from all FPI measurement between UTC times 00:57:04 and 01:00:33 (total of 1399). The velocity and temperature components are marked in differently colored unfilled histogram. All the reconstructions in this figure are performed at the highest allowable angular resolution ($L_{\rm{max, Nyq}} = 16$) for the MMS-FPI instrument grid.}
    \label{fig:MMS_collage}
\end{figure*}

We benchmark our SBR method on particle distributions measured by the FPI instrument onboard MMS spacecraft during the same interval as used in \cite{Servidio_etal_2017}. The results are shown in Fig.~\ref{fig:MMS_collage}. Panels (A1)-(A6) show the measurements as a function of energy shell. We present six energy shells from 584.16 eV through 2042.53 eV which contain the peak in particles distribution at time 2016-01-11/00:58:24 (80 seconds from the start of the interval). Measurement grids in $(\theta,\phi)$ which record null particle counts are left \textit{white}. Panels (B1)-(B6) show the reconstruction of the FPI distributions using Slepian basis on a polar cap. After investigating the FPI measurements during the data interval, we found that the distributions were dominantly within an $85^{\circ}$ polar cap. As mentioned before, the FPI measurement grid resolution limits us to an upper limit of $L_{\rm{max, Nyq}}=16$ in the wavenumber of our Slepian functions. A detailed demonstration of the energy-shell at E~=~584.15 eV is presented in Fig.~\ref{fig:Sleprec_demo_MMS} of Appendix~\ref{appendix: slepian_background}. In order to visually assess the reconstruction accuracy, we compare panel (A1) with (B1), (A2) with (B2) and so on. Clearly, the Slepian reconstruction captures all coherent structures in the data. This is because the Slepian basis is smooth and ensures that the reconstruction smoothly tapers off where the data has zero counts. The entire reconstruction process happens in the log-scale of distribution function. The associated colorbar has a minimum value of 0 in log scale which corresponds to unity in linear scale. 
% \sout{This is because we normalize the measured distribution function to have a minimum value of 1.0 before performing energy-shell fits to Slepian functions.} 
{\color{black}This is because for each energy shell, we normalize the distribution to have a minimum value of 1.0 before performing the fits to Slepian functions.}

Although the reconstructions happen entirely using the energy-shell measurements, one at a time, we present the more customary $V_x - V_y$ slice in a Cartesian coordinate system in panels (C1)-(C3). It should be noted that since the FPI data does not have a measurement grid at exactly $\theta = 0$ (the plane corresponding to the $V_z = 0$ slice). From comparing the slices at the nearest elevation angles $\theta = \pm 5.625^{\circ}$, we know that the distribution function looks very similar. So, we have chosen the $\theta = 5.625^{\circ}$ slice for all the three panels and refer to it as the $V_x - V_y$ plane. Panel (C1) shows the slice in the FPI data with zero count measurement grid left in \textit{white}. Panel (C2) shows the result of fitting this distribution with a Slepian basis. This is in agreement with the data in panel (C1). As also noted when comparing panels (A1)-(A6) with panels (B1)-(B6), only the apparently coherent structures are preserved in the reconstruction. The patchy single-grid (not contiguous) measurements are washed out because they fall below the Nyquist angular resolution of the FPI measurement. An advantage of using a basis parameterization of the distribution (while ensuring we do not go higher than the allowed Nyquist angular resolution) is that once we have the coefficients of the Slepian functions, say at a given energy shell, we can generate high-resolution maps of the distributions using higher resolution Slepian functions in $(\theta, \phi)$ with the same Slepian coefficients. Panel (C3) shows the high-resolution reconstruction using the Slepian coefficients found when generating the FPI resolution distribution in panel (C2).

Finally, the bottom panels in Fig.~\ref{fig:MMS_collage} compare the moments calculated from the FPI data and the moments calculated from our SBR method. To quantify the robustness of the reconstruction method, we take the ratio of the SBR to FPI moments (which for exact recovery will be 1). The bottom left panel of Fig.~\ref{fig:MMS_collage} shows the distribution of the density ratio. The most-probable ratio is \textcolor{black}{$0.983^{+0.0078}_{-0.0147}$}, where the superscript and subscript refer to $1\sigma$ denoting the 14th and 86th quantile, respectively. The bottom middle panel shows the ratio for velocity magnitude (solid blue distribution) and the vector components (colored outlines). We find agreement between the original and reconstructed velocity moments. Fig.~\ref{fig:MMS_collage} shows that the ratios for the velocity magnitude is (\textcolor{black}{$1.001^{+0.0029}_{-0.0032}$}), x-component is (\textcolor{black}{$1.001^{+0.0032}_{-0.0033}$}), y-component is (\textcolor{black}{$1.008^{+0.0091}_{-0.0052}$}), and z-component is ($\textcolor{black}{0.992^{+0.0137}_{-0.0179}}$). Finally, the bottom left plot shows the ratio for the trace of the temperature tensor (solid distribution) and the diagonal tensor components (colored outlines).

It is important to clarify that the moment calculations discussed here differ from the exact pipeline used to generate L2 moments supplied by the instrument team. We utilize our own numerical integration approach in order to make an \textit{apples to apples} comparison of our reconstructed moments. A standard Simpson's integration method is employed to determine both the FPI and SBR moments, which are defined on the instrument grid.

% There is a very strong linear correlation and the points fall largely on the $x=y$ line inclined at $45^{\circ}$ marked in \textit{dashed red}. This shows that our reconstruction conserves the structure as well as the overall plasma moments of FPI distributions.

\subsection{Application to SWA-PAS ion distributions} \label{sec: SOLO_REC}
\begin{figure}
    \centering
    \includegraphics[width=\linewidth]{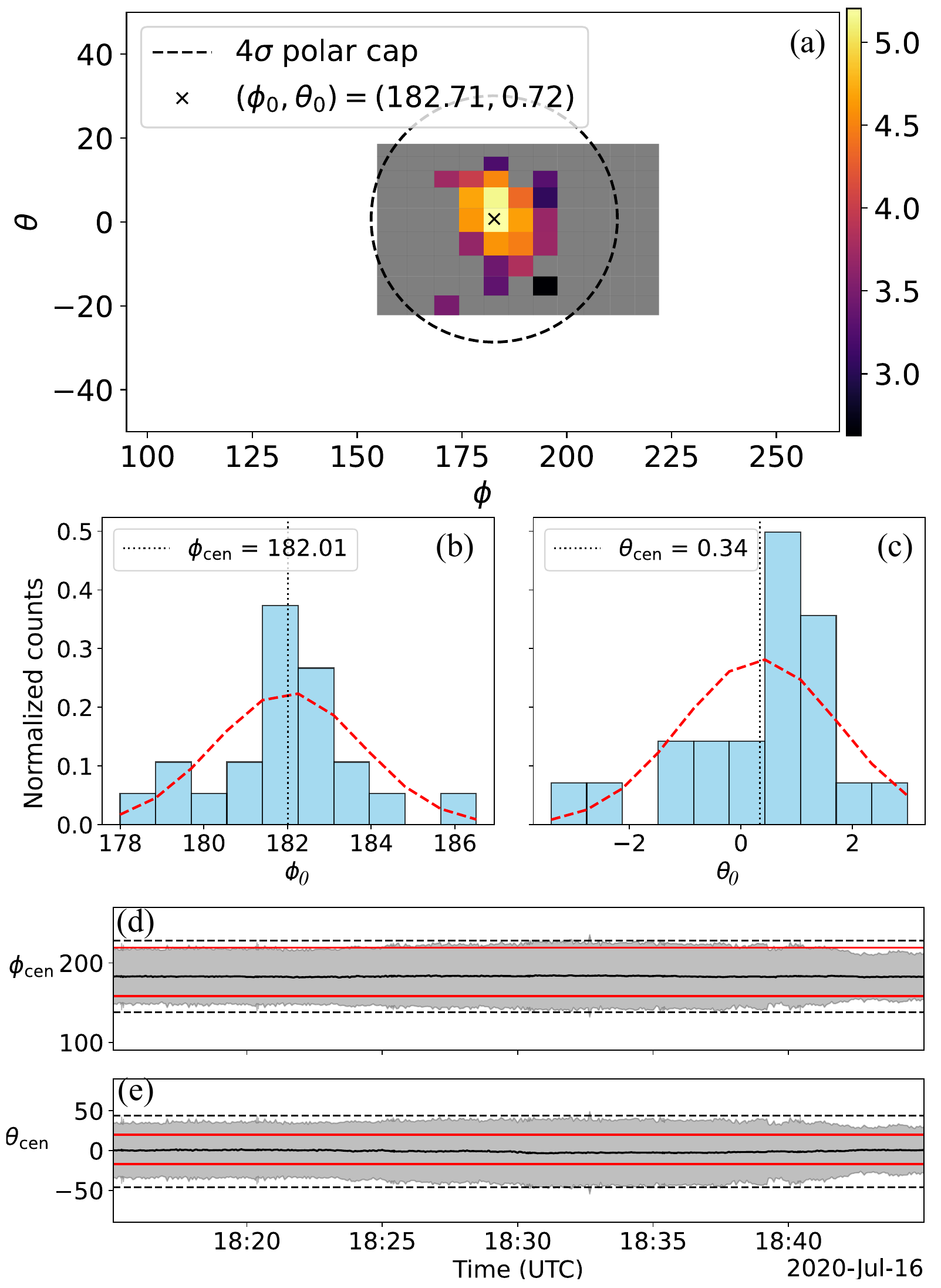}
    \caption{\textbf{Three-step method to find the Slepian polar cap domain.} Panel (a) shows the first step where we fit a 2D Gaussian in the $(\phi, \theta)$ plane to estimate the approximate centroid of the VDF $(\phi_{0}, \theta_0)$ for each energy shell. Panels (b) \& (c) \textcolor{black}{
    show normalized histograms of bin centers for the $\phi_{0}$ and $\theta_{0}$ centroid locations for all energy shells with sufficient data measurements.}
    % the normalized histogram for $\phi_0$ and $\theta_0$ calculated from all energy shells with sufficient counts. 
    The peak of the fitted Gaussian (\textit{red dashed} lines) is the effective centroid location $(\phi_{\rm{cen}}, \theta_{\rm{{cen}}})$ of the VDF across energy shells for a single timestamp. Panels (d) \& (e) show the variation of the centroid location and $4\sigma$ shading around it, where $\sigma$ is the maximum standard deviation (as a function of energy) of the 2D Gaussian fitted in Panel (a). The \textit{solid red} horizontal lines show the angular extent of the SWA instrument. The \textit{black dashed} horizontal lines show the extent of the polar cap we have chosen for the analysis presented in Fig.~\ref{fig:SolO_collage}.}
    \label{fig:capfit}
\end{figure}

\begin{figure*}
    \centering
    \includegraphics[width=\linewidth]{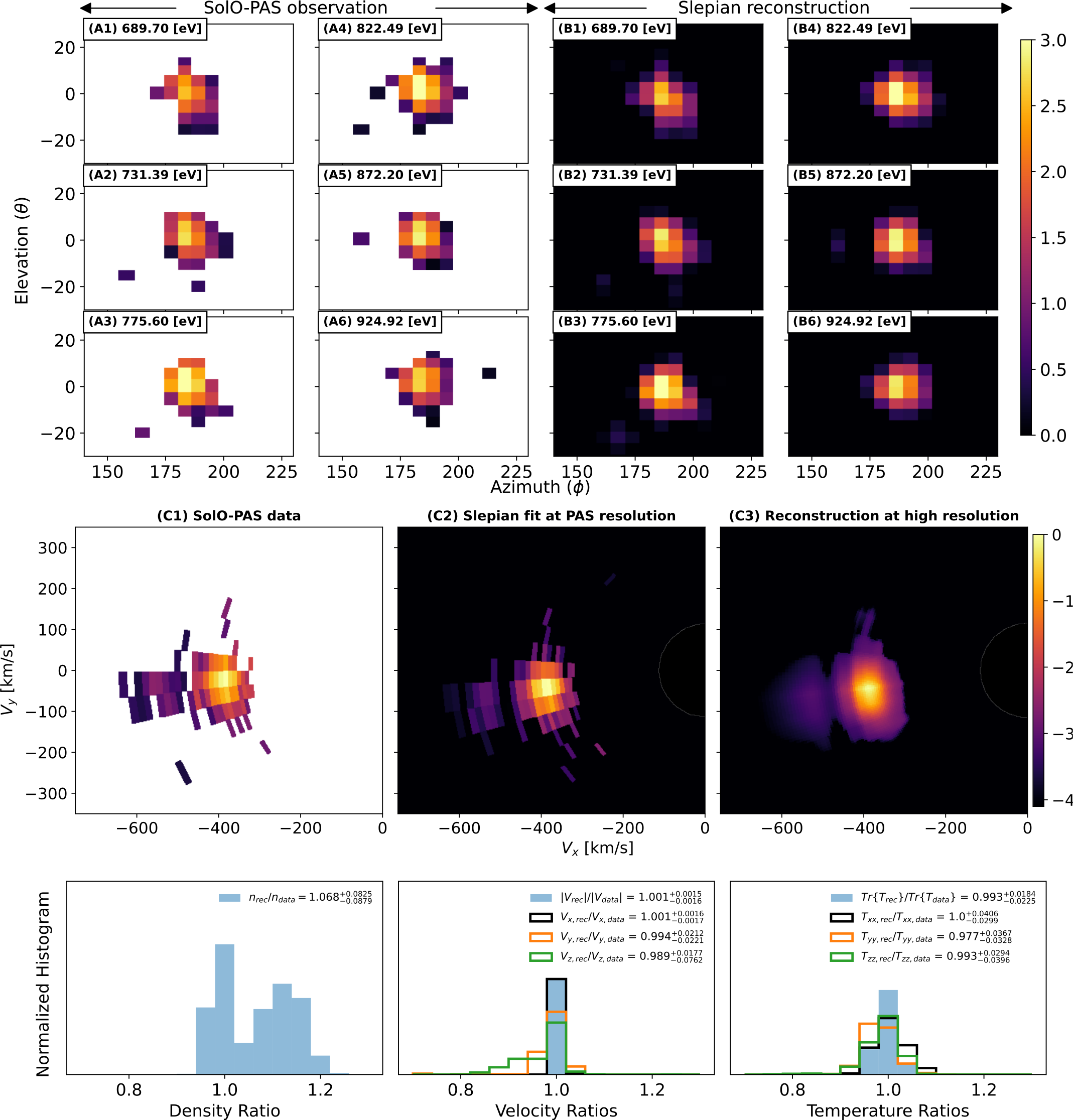}
    \caption{\textbf{SolO-PAS energy shell reconstruction plots.} Figures on the two left columns (A1)-(A6) show selected energy channels of SolO-PAS observation at time 2020-07-16/18:18:31 UTC. The corresponding energy shell values are written in the panel headers. The right two columns show the corresponding reconstructions in (B1)-(B6), respectively. Panels (C1)-(C3) shows the data and reconstructions on the closest Cartesian slice to the $V_x - V_y$ plane. (C1) shows the slice of SolO-PAS data at $\theta = 2.435^{\circ}$ off the $V_z = 0$ plane and (C2) shows the Slepian reconstruction at the same resolution as the PAS measurement. (C3) shows a high resolution reconstruction where we used a cubic B-spline interpolation in energy to ensure smoothness and a higher resolution Slepian function weighted by the same coefficients as inferred from the PAS resolution Slepian functions. The \textit{bottom} panel compares the density and velocity moments obtained from the data and reconstruction. Each histogram corresponds to the moment ratios from all PAS measurement between UTC times 18:15:03 and 18:44:59 (total of 612). The velocity and temperature components are marked in differently colored unfilled histogram. All the reconstructions in this figure are performed at the highest allowable angular resolution ($L_{\rm{max}} \simeq 28$) for the SolO-PAS instrument grid.}
    \label{fig:SolO_collage}
\end{figure*}

We use measurements from the Proton and Alpha particle Sensor (PAS) which is a part of the Solar Wind Plasma Analyzer \citep[SWA,][]{SWA} instrument suite onboard SolO for demonstrating the potential of using our SBR methodology in solar wind plasma with \textit{complete} field-of-view coverage. As in the case of MMS, this exercise also involves using basis reconstruction using Slepians on a polar cap but has quantitative differences by virtue of the highly localized distributions in $(\phi, \theta)$ velocity phase space. Consequently, defining the polar cap domain to be precisely around the location of the distribution is crucial in limiting the number of basis functions required to parameterize it. Again, this is one of the primary advantages of using the optimally localized Slepian functions as compared to basis functions that span all of $\mathbb{R}^2$ space, such as spherical harmonics.

For carrying out SBR on PAS measurements, we first need to find the polar cap domain in the velocity phase space over which the distributions exist. This exercise involves broadly three steps which are demonstrated in Fig.~\ref{fig:capfit}. Panel (a) shows the energy slice at 1870.53 eV in the $(\phi, \theta)$ velocity phase space. All available PAS measurement grids are shown in the rectangular grey patch. The colored grids show non-zero counts at that particular timestamp. This represents the PAS measurement. In the first step we fit a two-dimensional isotropic Gaussian to this measurement in the $(\theta, \phi)$ space. For each energy shell with substantial\footnote{Since the 2D Gaussian fitting optimize for 3 parameters, we need atleast three grids with non-zero measurement counts in an energy shell. It is also important to discard outlier energy channels where the measurements are nearly along one channel in $\theta$ or in $\phi$ which is insufficient to constrain the 2D Gaussian. For energy shell where either of the above are not a concern are termed energy shells with \textit{substantial} counts} counts, the peak location $(\phi_0, \theta_0)$ and the standard deviation $\sigma$ are optimized during this fitting process. For the energy shell shown in panel (A), the \textit{black} cross shows the peak location $(182.71^{\circ}, 0.72^{\circ})$ and the \textit{dashed black} circle shows the $4\sigma$ boundary. For each timestamp, we make a histogram for the fitted $\phi_0$ and $\theta_0$ across energy shells. This is done with the motivation of finding one effective centroid location $(\phi_{\rm{cen}}, \theta_{\rm{cen}})$ for all energy shells. Finding an effective centroid for each shell is essential because (i) the generation of Slepian functions for each energy shell is computationally expensive, (ii) for instances where the gyrotropy assumption is implemented a global axis of symmetry for the entire VDF needs to be identified, and, (iii) for analyzing multiple intervals (as we will later in this section), generating one Slepian basis requires us to find an all-encompassing polar cap for that period to speed up calculation across timestamps.

For each timestamp, we then build histograms as shown in panel (b) \& (c) to find an effective centroid location of the polar cap $(\phi_{\rm{cen}}, \theta_{\rm{cen}})$ for the distribution. We use the respective peak of the fitted Gaussian shown in \textit{red dashed} line as our centroid location for that timestamp. These centroid locations are shown as a function of time in panels (d) \& (e). It is notable that the centroid location of the distributions across the time interval we choose is stably localized inside the bounds of the instrument's angular extent (shown by the \textit{solid red} horizontal lines). The $4\sigma$ value across this interval is shown in the \textit{grey} patch around the centroid locations. Since we analyze the entire time period, we choose a polar cap $\sim 45^{\circ}$ in angular extent around the mean of the $(\phi_{\rm{cen}}, \theta_{\rm{cen}})$ for fitting each of the distributions across all energy shells at each time stamps in the entire time interval.

Figure~\ref{fig:SolO_collage} demonstrates the VDF reconstruction from measurements by the SWA-PAS onboard SolO. We use the same time interval as in \cite{Lavraud_etal_2021} spanning 18:15:03 UTC to 18:44:59 UTC on July 16, 2020. In panels (A1)-(A6) we show the measured distribution functions on the six energy shells containing the peak intensity. The Slepian reconstructions are shown in panels (B1)-(B6). As in Fig.~1, (A1) should be compared to (B1), (A2) to (B2), and so on and so forth. The normalization and colorscale are chosen in a similar fashion as described for the MMS-FPI demonstration. Panels (C1)-(C3) show the more conventional representation of VDFs in the $V_x - V_y$ phase space. (C1) shows the original PAS measurements with white patches representing a null count. (C2) shows the Slepian reconstruction at the same resolution as the PAS instrument grid and (C3) shows a super-resolved version of the reconstruction in (C2) using the same coefficients but a finer mesh for constructing the Slepian basis. Finally, in the bottom panel we show histograms of moment ratios (data vs. reconstruction). Excellent agreement is found for moments computed on the original measurements and the Slepian reconstruction. The most probable value for (a) density ratio is \textcolor{black}{$1.068^{+0.0825}_{-0.0879}$}, (b) velocity magnitude is \textcolor{black}{$1.001^{+0.0015}_{-0.0016}$}, (c) $V_x$ component is \textcolor{black}{$1.001^{+0.0016}_{-0.0017}$}, (d) $V_y$ component is \textcolor{black}{$0.994^{+0.0212}_{-0.0221}$}, (e) $V_z$ component is \textcolor{black}{$0.989^{+0.0177}_{-0.0762}$}, (f) trace of the temperature tensor is \textcolor{black}{$0.993^{+0.0184}_{-0.0225}$}, (g) $T_{xx}$ is \textcolor{black}{$1.0^{+0.0406}_{-0.0299}$}, (h) $T_{yy}$ is \textcolor{black}{$0.977^{+0.0367}_{-0.0328}$}, and (i) $T_{zz}$ is \textcolor{black}{$0.993^{+0.0294}_{-0.0396}$}. We want to note that all reconstructions for SolO-PAS is performed at the maximum resolvable Nyquist angular limit of $L_{\rm{max}} = 28$. However, since from Fig.~\ref{fig:capfit} we know that the PAS measurements are contained inside a $45^{\circ}$ polar cap, we only use the Slepian functions optimally concentrated inside this polar cap. This spatial concentration of the Slepian bases is implemented by truncating beyond a given number of functions (when arranged in decreasing order of localization within the polar cap). This upper limit of the number of bases to achieve the polar cap localization is called the Shannon number. The reader is directed to Appendix~\ref{appendix: slepian_background} for further details.

\color{black}
\subsection{Quantifying Slepian reconstruction fits}\label{sec:Ratios}
This section compares the measured MMS and SolO VDFs to their Slepian reconstructions to illustrate the ability to preserve fine-scale kinetic features. By virtue of using smooth Slepian basis functions, the SBR method facilitates performing numerical derivatives in velocity phase space. While we have demonstrated the moment preserving property of the Slepian fits in Sections.~\ref{sec: MMS_REC}~\&~\ref{sec: SOLO_REC}, this does not immediately imply phase-space structure preservation. Since this is of significance to maintain the essential characteristics of the distribution function (i.e., beams, tails, halos, etc), here we perform a direct comparison of the VDF measurements with the SBR reconstruction. 

\begin{figure*}
    \centering
    \includegraphics[width=\linewidth]{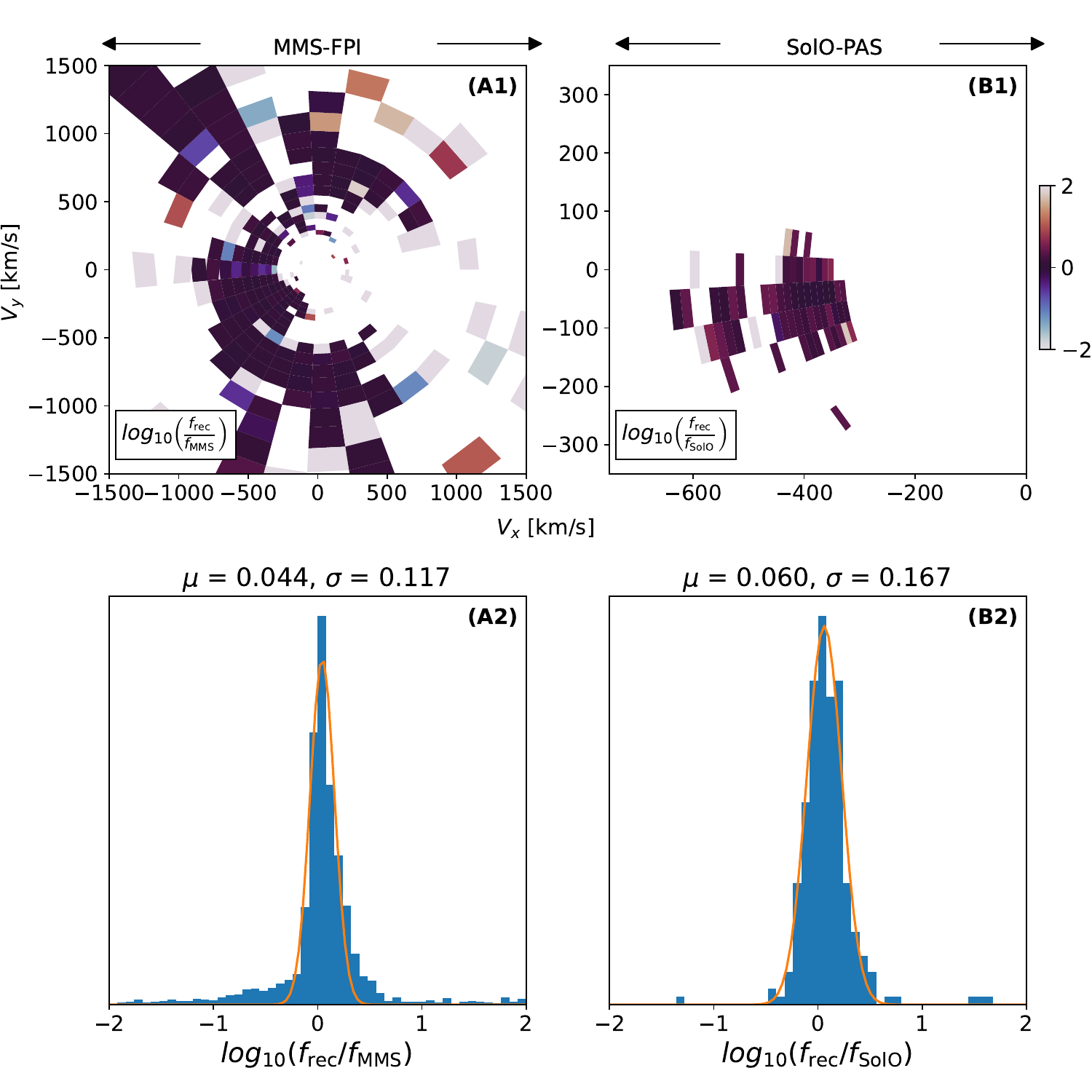}
    \caption{\color{black}\textbf{MMS-FPI and SolO-PAS measured vs. reconstructed distributions.} Figures on the left half panel quantify the goodness of reconstruction of the measured MMS-FPI observation. The right half panel shows the same for a SolO-PAS measurement. The selected MMS and SolO events are the same as those in Figs.~\ref{fig:MMS_collage} and \ref{fig:SolO_collage}, respectively. Panels (A1) \& (B1) shows the logarithm of the ratio of the reconstruction VDF to the measured VDF. Only the angular grids with finite count measurements are plotted. These $V_x - V_y$ planes are constructed from the slices of MMS-FPI and SolO-PAS on the closest Cartesian slice. Histograms showing a more concise quantification of the goodness of reconstruction, for entire 3D grid, is shown in panels (A2) \& (B2). The \textit{orange} line denotes a best Gaussian fit to the histogram. The resultant mean $\mu$ and standard deviation $\sigma$ are printed in the respective figure titles.}
    \label{fig:ratios}
\end{figure*}

To compare the SBR reconstructed VDFs ($f_{\mathrm{rec}}$) with those measured by MMS-FPI and SolO-PAS ($f_{\mathrm{obs}}$) we quantify the log relative error as $\rm{log}_{10}(f_{\rm{rec}} / f_{\rm{obs}})$.
% \begin{align}
%     \epsilon_r = \frac{f_{\mathrm{rec}} - f_{\mathrm{obs}}}{f_{\mathrm{obs}}}.
% \end{align}
Panels (A1) \& (B1) in Figure~\ref{fig:ratios} show the log relative error for the MMS while panels (A2) \& (B2) show the same for SolO. The chosen events are the same as those presented for MMS and SolO in Figs.~\ref{fig:MMS_collage} and \ref{fig:SolO_collage}, respectively. Comparing Fig.~\ref{fig:ratios}(A1) with Fig.~\ref{fig:MMS_collage}(C1) and Fig.~\ref{fig:ratios}(B1) with Fig.~\ref{fig:SolO_collage}(C1), we see that the highest intensity grids of the measured VDFs are in reliable agreement with the reconstruction. As expected, the largest offsets between data and reconstructions appear along the edges (or at low intensity grids) of the finite counts region, where the signal is predominantly influenced by Poisson noise. 
Given that Slepian reconstructions transition the distribution from the core to nearly zero in the periphery, these noisy regions manifest as more pronounced difference pixels in a direct comparison map in the $V_{x}-V_{y}$ plane. This is due to the inherent smoothening of the basis function representation, as several of the noisy, scattered points are smoothly connected with the core of the distribution in the SBR reconstruction. This demonstrates that our reconstructions offer a high-fidelity representation of the high signal-to-noise portion of the data and effectively eliminate spurious noise effects, which would otherwise render the unprocessed measurements inadequate for dispersion solvers such as ALPS.

The bottom panels (A2) \& (B2) in Fig.~\ref{fig:ratios} presents histograms of grid-wise log relative errors for the entire 3D distribution. The fitted Gaussians $\mathcal{N}(\mu, \sigma)$ to the histogram are shown in orange. Converting the inferred values of $\mu$ and $\sigma$ from log space to linear space, we see that on an average the reconstruction is $10^{0.044} \sim 1.1$ times the MMS-FPI measurement. Similarly for SWA-PAS, an average reconstruction is about $10^{0.06} \sim 1.14$ times the measurement. Moreover, less than 5\% finite measurements are over or underestimated by a factor larger than 1.73 for MMS-FPI and by a factor of 2.15 for SolO-PAS. Note that the timestamps chosen here are for the purpose of demonstration and these goodness of reconstruction quantification would vary slightly across time-stamps for each instrument.
\color{black}

\subsection{Ion VDF with Spherical Harmonics: Motivation for Slepian Functions}\label{sec:AppendixD}

\begin{figure*}
    \centering
    \includegraphics[width=\linewidth]{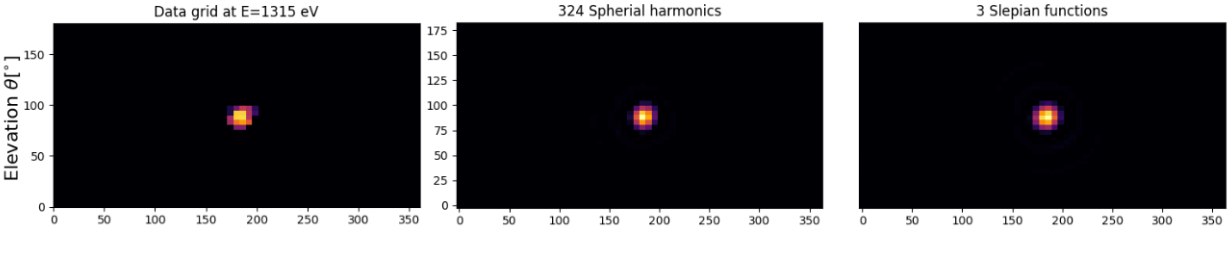} \\
    \includegraphics[width=\linewidth]{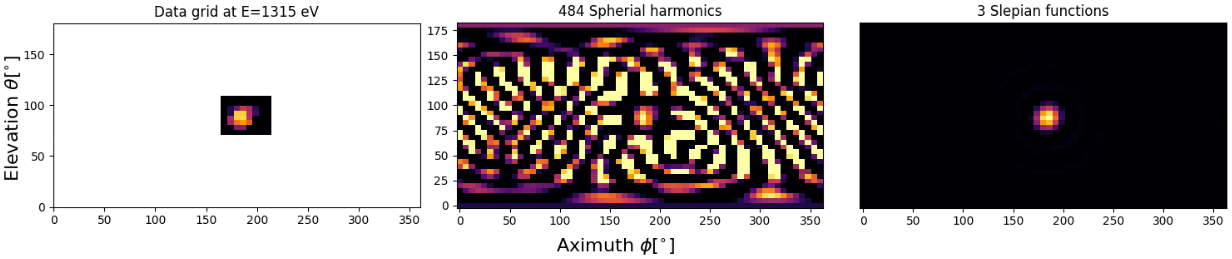}
    \caption{Top: Ion velocity distribution functions defined globally. Left: Original VDF pattern. Middle: Spherical harmonic reconstruction using 324 basis functions across the complete phase space. Right: Slepian reconstruction with three basis functions. Bottom: Ion velocity distribution function on the Solar Orbiter grid. The three panels above are repeated here, with the exception that the middle panel utilized 484 spherical functions harmonics.}
    \label{fig:SH_vs_SLEP}
\end{figure*}

For solar wind plasmas, the ion distribution functions are typically fast-moving and cold. Implying that the ions are well localized in phase space. Therefore, attempting to use globally defined orthogonal functions, such as spherical harmonics, to reconstruct ion VDFs is not optimal. Over the finite domain, the spherical harmonics no longer satisfy the orthogonality condition. 
Furthermore, hundreds of basis functions are needed to reproduce the original VDF pattern and fit cold plasma distributions successfully. 

Figure~\ref{fig:SH_vs_SLEP} illustrates the shortcomings of the globally defined Spherical Harmonic reconstruction when compared to the locally defined Slepian Basis for a cold solar wind plasma. The top panels of Fig.~\ref{fig:SH_vs_SLEP} assume a globally defined ion distribution function. A satisfactory reproduction of the original pattern requires 324 spherical harmonics. In contrast, only three Slepian functions can recreate the same detail, necessitating just one-hundredth of that count. To further emphasize the limitations of spherical harmonics in cold plasma, the bottom panel of Figure~\ref{fig:SH_vs_SLEP} replicates the analysis on an approximate Solar Orbiter grid. Here, 484 spherical harmonics were utilized to capture the original pattern but resulted in numerical artifacts outside the original domain. This brief demonstration highlights the effectiveness of locally defined orthogonal basis functions for ion distributions in the solar wind. This example underscores the inadequacy of globally defined basis functions for optimally reconstructing ion distribution functions in solar wind plasma conditions.

\section{Concluding Remarks} \label{sec:final_remarks}

We propose a new VDF reconstruction method that utilizes Slepian basis functions on a polar cap. For each energy shell in phase space, the Slepian basis is centered within a defined polar cap chosen to span the compact ion VDF. This ensures efficient parameterization of ion VDFs in both the hot magnetosheath and the cold solar wind, only requiring an optimal number of basis functions. 
Our method is an improvement over the spherical harmonic-based reconstruction methods for ion distributions as the Slepian basis functions are localized in space. Furthermore, our SBR method restricts the growth of numerical artifacts outside the instrument grid (Sec.~\ref{sec:AppendixD} and Fig.~\ref{fig:SH_vs_SLEP}), providing a more compact representation of the ion VDFs.
% This is an improvement on the spherical harmonics based reconstruction approach which require a much larger number of basis functions to achieve the required localization (especially for cold plasma cases). 
% Unlike spherical harmonics, our SBR method restricts the growth of numerical artifacts outside the instrument grid where localized solar wind VDFs are measured (see Sec.~\ref{sec:AppendixD} and Fig.~\ref{fig:SH_vs_SLEP}).

\begin{figure}
    \centering
    \includegraphics[width=\linewidth]{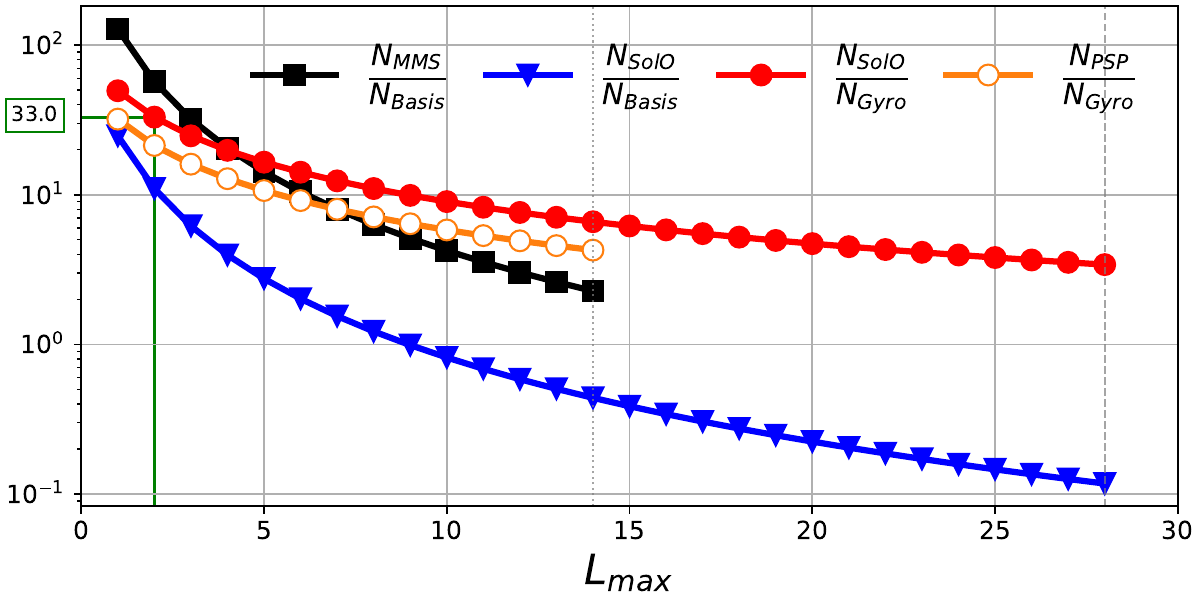}
    \caption{\textbf{Data points versus number of basis functions required.} This figure shows the ratio of the total number of angular grid points for each instrument as compared to the number of basis coefficients at different $L_{\rm{max}}$. The values in the y-axis is representative of the reduction in floating point numbers needed to represent the VDF. A larger value on the y-axis represents more efficient data storage. The maximum $L_{\rm{max}}$ we plot for each instrument is governed by the Nyquist limit (based on the angular grid resolution). This is shown by the vertical dotted line for MMS and PSP and by the dashed line for SolO.}
    \label{fig:ndata_vs_lmax}
\end{figure}

% When comparing Slepians with spherical harmonics (a family of globally defined orthogonal basis functions), a drastically fewer number of bases are needed for high-fidelity reconstructions.
We quantify the robustness of our SBR method by demonstrating that the reconstructions preserve plasma moments regardless of thermal extent while simultaneously capturing fine-scale complexities crucial for plasma kinetics. Given the drastically fewer bases required to generate high-fidelity reconstructions, the Slepian basis coefficients serve as a reduced representation of the measured VDFs. The ratio of the total number of data points per energy shell to the number of basis functions used over the $(\theta, \phi)$-grid 
% to parameterize the distribution in $(\theta, \phi)$ 
quantifies the efficiency of this reduction. We shall refer to this in the following discussions as the data compression factor. Fig.~\ref{fig:ndata_vs_lmax} shows the data compression ratio for different spacecraft measurements as a function of $L_{\rm{max}}$. The black line with square markers corresponds to potential data compression for MMS-FPI when using all Slepian bases up to angular degree $L_{\rm{max}}$, making $N_{basis}=(L_{\rm{max}}+1)^2$. The FPI instrument spans nearly s $4\pi$ phase space distribution with an angular resolution of $11.25^{o}$ for 512 measurements per 32 energy shells. The SBR method results in a $\sim2.3$ times data compression at the Nyquist limit ($L_{\rm{max, Nyq}} = 14$, vertical dotted line). 
The choice of $L_{\rm{max}}$ depends on the granularity of the distribution function. 
% For a smooth VDF in $(\theta,\phi)$, a lower choice of $L_{\rm{max}}$ may be sufficient --- further increasing the data compression ratio beyond 2.3. 
For a VDF without fine-scale structures, a lower choice of $L_{\rm{max}}$ may be sufficient to maintain the bulk plasma moments, increasing the data compression ratio beyond 2.3.
The blue line with triangle markers shows a similar calculation of the data compression factor for SolO-PAS. Since PAS has an angular resolution $\sim6^{o}$, it has a significantly large Nyquist limit of $L_{\rm{max,Nyq}} = 28$, resulting in a much larger number of allowable basis functions. However, by virtue of its targeted window in angular space spanning only 11 azimuths and 9 elevations, the ratio $N_{data} / (L_{\rm{max}}+1)^2$ is not nearly as efficient as MMS. Since SolO measures a localized portion of angular phase space, the bases beyond the Shannon number are truncated to ensure localization (see Appendix~\ref{appendix: slepian_background}), further increasing the compression ratio.
% This would reduce the $N_{basis}$ and increases the data compression ratio. 
% The data compression factor is further boosted when considering the fact that the solar wind is dominantly gyrotropic.
The data compression factor is further boosted when considering the solar wind to be dominantly gyrotropic.

In a low-$\beta$ plasma, where $\beta = nk_bT/(B^{2}/2\mu_0) \leq 1$, the strong background magnetic field serves as a gyrotropic axis of symmetry. Since the solar wind is primarily gyrotropic, this renders an inherent symmetry to the VDF reconstruction setup. We aim to leverage this symmetry using only the axisymmetric Slepian basis functions anchored about the magnetic field axis. This drastically reduces the total number of basis functions from $(L_{\rm{max}}+1)^2$ to $(L_{\rm{max}}+1)$. In Fig.~\ref{fig:ndata_vs_lmax}, we show the data compression factor when considering only $N_{gyro}$ number of bases with the circular markers, red for SolO-PAS and orange for PSP-SPAN. Again, this is still a conservative estimate since we will truncate the Slepian basis beyond the Shannon number (to choose only the locally confined functions). This will further increase the data compression ratio beyond what is shown in the figure.

For the readers' convenience, we list our key conclusions below:
\begin{enumerate}[nosep]
    \item The Slepian basis is an optimal choice when reconstructing localized, cold plasma distributions in the solar wind. 
    % For hot non-thermal plasma distributions as observed inside planetary magnetospheres, spherical harmonics are an equally viable basis.
    \item In cases such as SolO-PAS and PSP-SPAN where the field of view does not span all $4\pi$ angular degrees, the concentration of Slepian functions within the domain of interest (given by the instrument grid) makes it an optimal choice as a basis function for reconstructions. The concentration of the bases within the observable instrument phase space 
    % This is because the bases concentration within instrument observable phase space 
    makes the inverse problem well-posed. This is not true for other bases which are global in nature, such as spherical harmonics.
    \item Slepian reconstructions maintain phase space complexities while preserving plasma moments. This is an improvement over a bi-Maxwellian representation, which may not capture all non-thermal features.
    % that inform essential macroscopic plasma properties. 
    % This is true for a general non-thermal distribution functions, which are otherwise poorly represented when using simple bi-Maxwellian distributions.
    \item The choice of $L_{\rm{max}}$, which is connected to the number of basis functions required, depends on angular grid resolution and granularity of the VDF in angular phase space. As shown in Fig.~\ref{fig:ndata_vs_lmax}, the SBR method can be a useful data compression technique. This is especially true in solar wind VDFs, which are dominantly gyrotropic about the magnetic field.
    \item Reconstructions performed with a chosen number of basis functions (bounded by the Nyquist limit in $L_{\rm{max,Nyq}}$) may be super-resolved onto any finely sampled angular grid. Hence, the SBR method alleviates the need for a traditional interpolation in the angular space of $(\theta, \phi)$. 
    \item The above outlined computational framework of applying the SBR method to MMS and SolO data has been developed under a Python and Matlab-based package, which we call \texttt{vdfit}\footnote{Interested readers are encouraged to get in touch with the authors if they want to use \texttt{vdfit} in future studies.}. Future work will extend this framework to include solar wind ESA instruments PSP-SPAN and Helioswarm.
\end{enumerate}

% It is important to note that implementing the SBR method into an instrument design is outside the scope of this paper. For future applications, the Slepian basis functions can be defined on the exact spacecraft grids, allowing for a simple one-to-one calculation of the Slepian coefficients. 

The bulk of the machinery developed in this study was done in preparation for application to current and future solar wind missions such as PSP and Helioswarm. We first plan to apply our method to SPAN-i onboard PSP. The heat shield restriction in SPAN often renders measurements of the VDF incomplete \citep{Livi_etal_2022}. Considering the gyrotropic assumption and using the minimal number of Slepian bases would drastically improve the condition number of the inverse problem for recovering the full VDF (from partial observations). Cases where the part of the core or beam is blocked due to heat shield restrictions can potentially be recovered. This would enable calculations of complete moments (as opposed to partial moments) where the underlying model VDF is not an oversimplified bi-Maxwellian.  Although, the feasibility would also depend on the fraction of VDF being blocked off and the robustness of the estimated magnetic field direction. In otherwise occulted cases, potential recovery of the core and beam will open avenues to robustly infer temperature anisotropy and lend insight into plasma heating processes. In the case of Helioswarm, eight nodes with Faraday cups will be complemented by measurements from an electrostatic analyzer onboard the central hub \citep{klein2023helioswarm}. Recovering 3D (or 2D gyrotropic) VDFs on the nodes would provide a valuable complementary measurement to addressing critical questions in plasma turbulence. 
% This is an interesting inverse problem in its own merit. 
Our SBR method proposes a promising candidate basis function (model parameter for the inverse problem) to model VDFs that are not in equilibrium. 
% which are essential to study turbulence \citep{?}. 
Each Helioswarm node (1D VDF) has a low number of measurements, making the inverse problem under-determined. Gyrotropic Slepians on a polar cap form a minimal basis set, which reduces the number of free parameters and enables better conditioning of the inverse problem.
% The fact that gyrotropic Slepians on a polar cap represent the minimal set of basis would be especially crucial in under-determined inverse problems, like the one we will have at the nodes of Helioswarm, where the number of data measurements (1D VDF) is low. 
Looking ahead, future mission concepts hope to reduce the angular and temporal resolution of electrostatic analyzers \citep[][]{De_Marco_2016, Wilson_2022}, significantly increasing the data counts and making the problem of inferring complex VDFs better conditioned.

Observations of ion VDFs near current sheet crossings give rise to various thermal and non-thermal ion distributions. Observations, near reconnection events within the current sheet, show multiple plasma populations \citep{Lavraud_etal_2021} and additional highly diffusive beams \citep{Verniero_2020}. The SBR method is agnostic to the plasma species and free from convergence issues arising from ad-hoc choices of initial guess parameters. Therefore, we can fit any number of plasma species observed by the ESA. However, the SBR method in its current form cannot distinguish the multiple plasma populations from a singular proton VDF, as we observe in the SolO-PAS reconstructions Fig.~\ref{fig:SolO_collage}. This is where novel machine learning and image processing techniques can be employed to identify and segregate multiple plasma populations \citep{De_Marco_2023}. Potential cross-over of these techniques with SBR opens promising avenues to quantify the energy transport.

Considering the SBR method's ability to effectively fit cold and hot non-equilibrium plasma distributions, a natural next step in this research is to analyze the linear dispersion relation of the reconstructed VDFs. To accomplish this, we plan to explore the linear modes using the Arbitrary Linear Plasma Solver (ALPS) 
\citep{Verscharen_ALPS_2018}. Previous studies have shown that deviations from bi-Maxwellian distributions affect the expected dispersion relations
\citep{Walters_2023}, suggesting a complicated interplay between changes in linear modes and local thermodynamic equilibrium. Additionally, we will investigate proposed kinetic mechanisms, such as the phase-space cascade 
\citep{Servidio_etal_2017, Wu_2023, Nastac_2024}, utilizing the novel reconstructions from the SBR framework. Given that reconstructions from SBR generate distributions that preserve plasma moments and enable super-resolution, we can avoid the constraints posed by various interpolation techniques \citep{terres2023energy}. By integrating these innovative tools with wave-particle correlation methods, numerical solvers, and phase space entropy cascades, we can now investigate kinetic processes in unprecedented detail from in-situ observations.

\begin{acknowledgements}
    SBD and MT were supported by the Parker Solar Probe mission SWEAP investigation under NASA contract NNN06AA01C. SBD and MT are particularly grateful to Frederik J. Simons for numerous discussions and the MATLAB software used in the generation of the Slepian functions. {\color{black} These may be found in the publicly available packages releazed in Zenodo as \href{http://dx.doi.org/10.5281/zenodo.592782}{\texttt{slepian\_alpha}} and \href{http://dx.doi.org/10.5281/zenodo.3676147}{\texttt{slepian\_foxtrot}}}. Additionally, the authors thank Michael L. Stevens for their guidance and detailed feedback throughout this study and on the manuscript. We acknowledge Kristoff Paulson for critical input and detailed discussion throughout our study. 
    % This research was funded by the grant number XXXXXXXXXX. We grateful to Frederik J. Simons for the MATLAB software to generate Slepian functions and for numerous discussions along the way. We thank Michael L. Stevens for detailed valuable feedback of the manuscript. We acknowledge the Kristoff Paulson for critical feedbacks during the duration of our study.
\end{acknowledgements}

%\bibliography{references}{}
%\bibliographystyle{aasjournal}

\appendix
\section{Mathematical outline of reconstruction using Slepian functions on a polar cap} \label{appendix: slepian_background}

\begin{figure}[H]
    \centering
    \includegraphics[width=\linewidth]{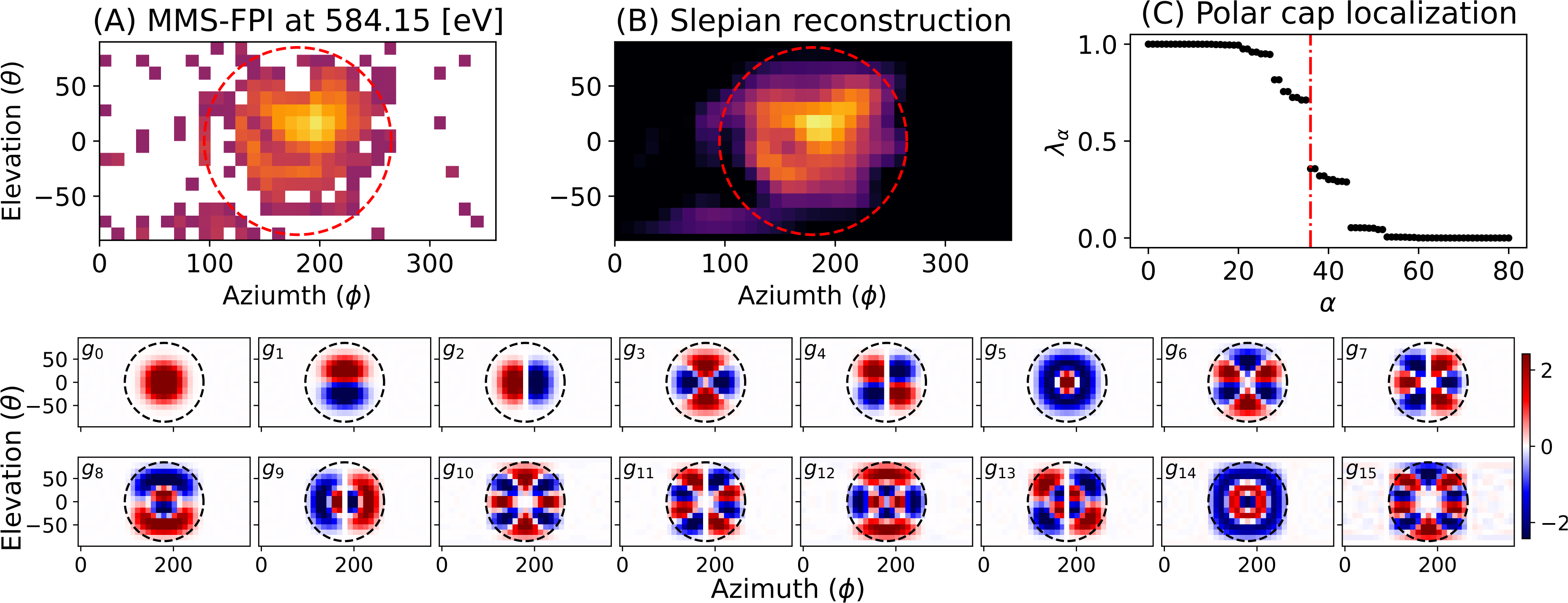}
    \caption{\textbf{Decomposing distribution function using Slepians on a polar cap.} Panel (A) shows the MMS-FPI distribution function at E = 584.15 eV with zero count measurement grid åre left \textit{white}. Panel (B) shows the corresponding Slepian reconstruction using basis functions generated on a polar cap of extent $85^{\circ}$ (demarcated with the \textit{red dashed} line. The maximum angular degree is chosen to be $L_{\rm{max}} = 8$. Panel (C) shows the localization value, quantified by the eigenvalue corresponding to each Slepian function $\lambda_{\alpha}$. The Slepian basis is arranged in descending order of localization, with $\lambda_{\alpha} = 1$ indicating a basis that is completely localized within the polar cap (see Eqn.~[\ref{eqn: lambda_alpha}]). The vertical \textit{red dashed} line in Panel (C) shows the Shannon number $\sim 36$, before which the basis functions are dominantly confined within the polar cap. The bottom panels show the first 16 Slepian functions $g_{\alpha}(\theta,\phi)$ used to perform the above decomposition.}
    \label{fig:Sleprec_demo_MMS}
\end{figure}

This section outlines the formalism of Slepian functions on a polar cap for the ease of the readers' reference. The tools for Slepian representation implemented here were first introduced in \cite{Slepian_polar_caps}. Since, for this study, we apply this method to complete field-of-view instruments (MMS and SolO), we shall represent the distribution function on polar caps in elevation ($\theta$) and azimuth ($\phi$) space.
% the function we want to decompose into Slepians on polar caps in the elevation $\theta$ and azimuth $\phi$ space.
For a given energy E, determined by the voltage applied across ESA deflector plates, the velocity distribution function can be written as $f_E(\theta,\phi)$. We want to parameterize this 2D distribution function on the surface of the sphere in the basis of polar cap Slepians $\mathbf{G}(\theta,\phi)$. For a given maximum angular degree $L_{\rm{max}}$, the Slepian basis $\mathbf{G}(\theta,\phi) = (.... , g_{\alpha}, ...)$ where $\alpha \in [0, (L_{\rm{max}+1})^2]$. {\color{black} Note that these space-concentrated basis functions on a polar cap are numerically computed by solving an eigenvalue problem as detailed in Sections~3~\&~4 of \cite{Slepian_polar_caps}}. It is conventional to arrange $\alpha$ is decreasing order of localization $\lambda_{\alpha}$ of $g_{\alpha}(\theta,\phi)$ within the domain $\mathcal{R}$ on a unit sphere $\Omega$, where
\begin{equation} \label{eqn: lambda_alpha}
    \lambda_{\alpha} = \frac{\int_{\mathcal{R}} g^2_{\alpha}(\theta, \phi) \sin{\theta} \, \rm{d}\theta \, \rm{d}\phi}{\int_{\Omega} g^2_{\alpha}(\theta, \phi) \sin{\theta} \, \rm{d}\theta \, \rm{d}\phi} \, .
\end{equation}

Therefore, in order to define space-concentrated and band-limited Slepian functions, we need to specify a domain in (phase) space and a bandwidth interval $0 \leq \ell \leq L_{\rm{max}}$. While the phase space domain $\mathcal{R}$ is inspired by the physical extent of $f_E$ in $(\theta, \phi)$ phase space, the maximum wavenumber $L_{\rm{max}}$ is decided by considering the degree of coarseness we want to resolve the distribution as well as the resolution limit of the observing instrument. 
Particle distributions measured by MMS, typically warm and agyrotropic, lack a well-defined symmetry axis and cover a large angular extent of $(\theta, \phi)$-space. 
% For particle distributions measured by MMS are typically warm and agyrotropic. Consequently they are considerably spread out in $(\theta, \phi)$ and lack a well-defined symmetry axis. 
Nevertheless, the distributions are dominantly contained within a portion on the $(\theta,\phi)$ spherical shell. For the MMS interval investigated in this paper, the dominant part of the distributions span nearly a full hemisphere. We specify the spatial extent of the polar cap (the Slepian domain in phase-space) to be $85^{\circ}$ around the centroid of the distribution function. For SolO, the particle distributions are much more narrowly confined and are dominantly gyrotropic. As mentioned in Sec.~\ref{sec:slepian_recon}, we use a spectral bandlimit of $L_{\rm{max}} = 180 / \Delta \theta$ which is the Nyquist wavenumber. This limit is enforced for each spacecraft measurement to ensure we do not over-fit the measurement and avoid introducing features post-fitting that exceed the instrument resolution. 

For a wavenumber given by $L_{\rm{max}}$, a distribution $f_E(\theta, \phi)$ can be decomposed in the basis of $\mathbf{G}(\theta,\phi)$ as
\begin{equation}
    f_E(\theta,\phi) = \sum_{\alpha=0}^{(L_{\rm{max}}+1)^2} c^{\alpha}_E \, g_{\alpha}(\theta,\phi) \, ,
\end{equation}
where, $\mathbf{G}(\theta, \phi) = (g_0(\theta,\phi), ..., g_{(L_{\rm{max}}+1)^2}(\theta,\phi))$. The first 16 (out of a total of 81) Slepian functions on a polar cap of angular extent $85^{\circ}$ for $L_{\rm{max}} = 8$, arranged in descending order of $\lambda_{\alpha}$, is shown in Fig.~\ref{fig:Sleprec_demo_MMS}. The Slepian decomposition coefficients $c_E^{\alpha}$ collectively capture the information of the distribution in each energy shell E, such that

\begin{equation} 
\left(\begin{array}{ccccc}
 g_0(\theta_0, \phi_0) & ... & g_i(\theta_0, \phi_0) & ... & g_N(\theta_0, \phi_0)\\
 \vdots  & &  \vdots & & \vdots \\
 g_0(\theta_j, \phi_j) & ... & g_i(\theta_j, \phi_j) & ... & g_N(\theta_j, \phi_j)\\
 \vdots  & &  \vdots & & \vdots \\
 g_0(\theta_M, \phi_M) & ... & g_i(\theta_M, \phi_M) & ... & g_N(\theta_M, \phi_M)\\
\end{array}\right)
\begin{pmatrix}
c^0_E \\
\vdots \\
c^i_E \\
\vdots \\
c^N_E
\end{pmatrix}=
\begin{pmatrix}
f_{E}(\theta_0, \phi_0) \\
\vdots \\
f_{E}(\theta_j, \phi_j) \\
\vdots \\
f_{E}(\theta_M, \phi_M) \\
\end{pmatrix}.
\end{equation}
If we choose to write the above equation as $\mathbf{G} \cdot \mathbf{C}_E = \mathbf{f}_E$, then the optimization of the coefficients is obtained from the standard Moore-Penrose damped pseudo-inverse
\begin{equation}
    \mathbf{C}_E = \left(\mathbf{G}^T \cdot \mathbf{G} + \mu \mathbf{I} \right)^{-1} \cdot \mathbf{G}^T \cdot \mathbf{f}_E \, ,
\end{equation}
where $\mu$ is the damping coefficient used to regularize an ill-conditioned inverse problem. Whether or not we need a non-zero $\mu$ depends on the coverage of data $f_E$ in $(\theta, \phi)$.

\begin{figure}[H]
    \centering
    \includegraphics[width=\linewidth]{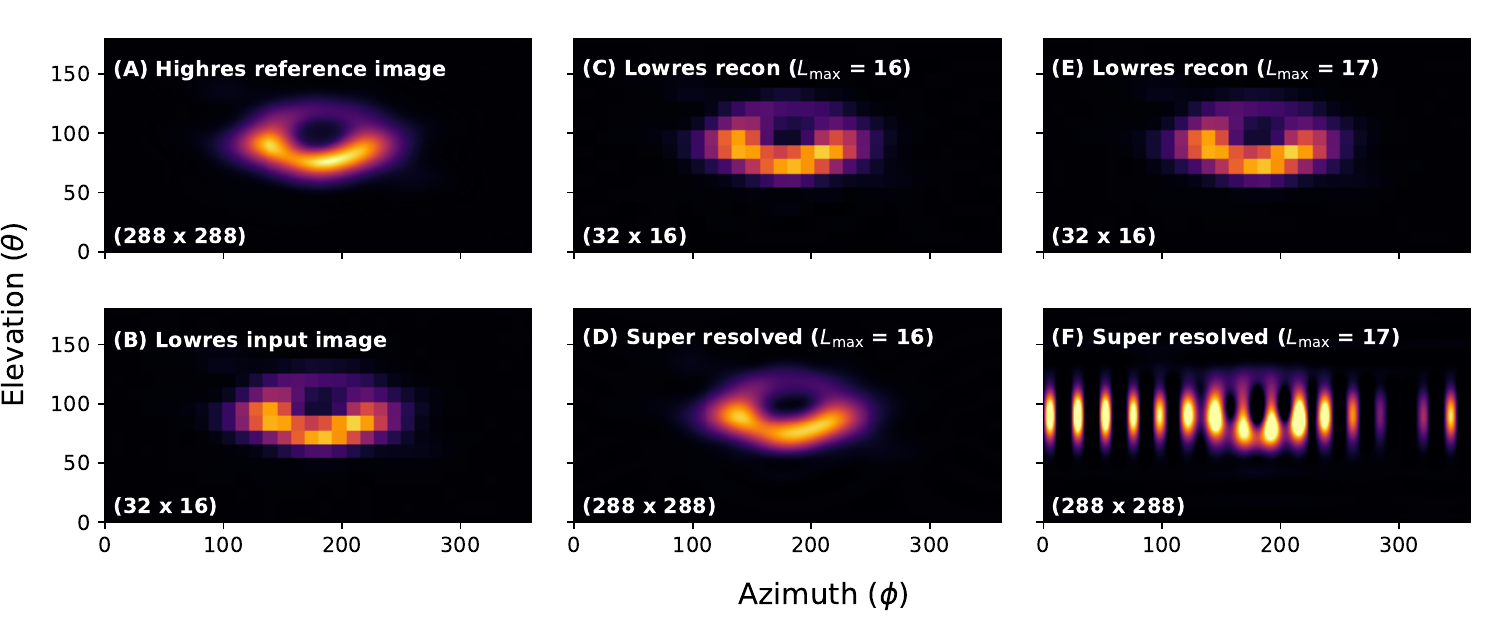}
    \caption{\textbf{Demonstration of maximum Nyquist limit on angular degree.} Panel (A) shows a high resolution reference image we use for the purpose of this demonstration. There are a total of (288 x 288) grids in $(\phi, \theta)$ space. In panel (B) we have used a simple linear interpolation to reduce the number of grids to (32 x 16) --- same as the MMS resolution. This is our \textit{input} image we use for the rest of the demonstration. In panel (C) we show the result of reconstructing the input image using basis functions truncated at $L_{\rm{max}} = 16$ which is the Nyquist limit for this grid resolution. Using the coefficients obtained when decomposing the input image in spherical harmonics in low resolution (32 x 16) grid, we carry out a \textit{super-resolution} by reconstructing using high resolution spherical harmonics. This is shown in panel (D). We carry out the same exercise in panels (E) and (F) by using $L_{\rm{max}} = 17$ which is one more than the Nyquist limit. Panel (E) is the reconstruction at low resolution and panel (F) is the corresponding super resolution using the coefficients obtained for reconstructing panel (E) but weighting high resolution spherical harmonics. The original reference image used here is the observation of the super massive based in the M87 galaxy observed by The Event Horizon Telescope in 2019.}
    \label{fig:Nyquist_demo}
\end{figure}

It is noteworthy to highlight the importance of not exceeding the maximum allowable angular degree in $L_{\rm{max, Nyq}}$ which is connected to the Nyquist limit imposed by the grid resolution. In our SBR implementation, we automatically impose an upper limit of $\rm{max}(180/\Delta \phi, 180 / \Delta \theta)$. For MMS, the Nyquist limit of $L_{\rm{max, Nyq}} = 16$. In Fig.~\ref{fig:Nyquist_demo} we present a simple demonstration of the power of super-resolution so long as we do not exceed $l_{\rm{max, Nyq}}$. For this demonstration we use a reference image --- the observation of the super massive based in the M87 galaxy observed by The Event Horizon Telescope in 2019. The high resolution image in (288 x 288) grids in $(\phi, \theta)$ space is shown in panel (A). This is interpolated using \texttt{scipy} to an instrument-resolution (32 x 16) grid to match MMS angular bin sizes. This reduces the maximum resolvable wavenumber in angular space to $L_{\rm{max, Nyq}}$. Next, we use this instrument-resolution image and decompose it into spherical harmonics generated on a (32 x 16) grid with maximum angular degree $L_{\rm{max, Nyq}} = 16$. The result of the reconstruction using the fitted coefficients is shown in panel (C). If we then use these fitted coefficients to construct a high resolution (288 x 288) image by weighing the high-resolution spherical harmonics, we see the reference image in panel (A) is very well recovered in panel (D). Finally, panel (E) shows the reconstruction at instrument-resolution when using $L_{\rm{max}} = L_{\rm{max, Nyq}} + 1$. At low resolution the reconstruction seems to be preserved on comparing panel (E) to panel (B), we see that the instrument-resolution reconstruction (using the low resolution fitting coefficients) is very different from the reference image in panel (A). This shows that exceeding the Nyquist limit when reconstructing distributions cannot be used for super-resolution since we would be introducing artifacts by virtue of over-fitting the measurements (by using basis functions that fluctuate at wavenumbers larger than that can be captured by the instrument angular resolution). Throughout this study, we ensure that this upper limit in angular resolution $L_{\rm{max, Nyq}}$ is enforced.

\end{document}